\documentclass[prc,twocolumn,superscriptaddress,showpacs,amssymb,amsmath,amsfonts,aps]{revtex4}
\usepackage[usenames]{color}
\usepackage{epsfig}
\usepackage{amssymb}
\usepackage{multirow}
\usepackage{longtable}
\usepackage{xcolor}
\newcommand{\be}{\begin{equation}}
\newcommand{\ee}{\end{equation}}
\newcommand{\bee}{\begin{eqnarray}}
\newcommand{\eee}{\end{eqnarray}}

\definecolor{grey}{rgb}{0.9,0.9,0.9}
\definecolor{black}{rgb}{0,0,0}

\def \irbaddress{Rudjer Bo\v{s}kovi\'{c} Institute, Bijeni\v{c}ka cesta 54, P.O. Box 180, 10002 Zagreb, Croatia}
\def \GWUSAIDaddress{Data Analysis Center at the Institute for Nuclear Studies, Department
of Physics, The George Washington University, Washington, D.C. 20052}
\def \mainzaddress{Institut f\"{u}r Kernphysik, Universit\"{a}t Mainz, D-55099 Mainz, Germany}
\def \jlabaddress{Jefferson Lab, 12000 Jefferson Avenue, Newport News, Virginia 23606, USA}
\def \bonnaddress{Helmholtz-Institut f\"{u}r Strahlen- und Kernphysik, Universit\"{a}t Bonn, Germany \\ Petersburg Nuclear Physics Institute, Gatchina, Russia}
\def \Osakaaddress{Department of Physics, Osaka University, Toyonaka, Osaka 560-0043, Japan}
\def \Osaka1address{Research Center for Nuclear Physics,
Osaka University, Ibaraki, Osaka 567-0047, Japan}
\def \bonnaddressdeborah{Helmholtz-Institut f\"ur Strahlen- und Kernphysik (Theorie) and Bethe Center for Theoretical
Physics,  Universit\"at Bonn, Nu\ss allee 14-16, 53115 Bonn, Germany}

\begin{document}

\title{ \Large Status of some P-wave baryon resonances \\ and \vspace*{1.cm} importance of inelastic channels }

\author{Volker D. Burkert}
\affiliation{\jlabaddress}
\author{Hiroyuki Kamano}
\affiliation{\Osaka1address}
\author{Eberhard Klempt}
\affiliation{\bonnaddress}
\author{Deborah R\"{o}nchen}
\affiliation{\bonnaddressdeborah}
\author{Andrey V. Sarantsev}
\affiliation{\bonnaddress}
\author{Toru Sato}
\affiliation{\Osakaaddress}
\author{Alfred \v{S}varc}
\affiliation{\irbaddress} 
\author{Lothar Tiator}
\affiliation{\mainzaddress}
\author{Ron L. Workman}
\affiliation{\GWUSAIDaddress}

\date{\today}

\begin{abstract}
\vspace*{1.cm}
\centerline{\textbf{Abstract}}
\vspace*{0.7cm}
We analyze the current status of three  P-wave baryon states $N(1710){1/ 2}^+$, $N(1900){3/2}^+$, 
and $\Delta(1600){3/2}^+$ as given in the Review of Particles Physics (RPP).  
Since the evidence for a particle's existence is linked to its RPP "star" rating, we discuss its 
subjective present definition. We also present the accumulating evidence supporting these states 
and give our new "star" rating 
recommendations. 

\end{abstract}

\maketitle

\section{Introduction}
In recent years, the field of light-flavor baryon spectroscopy has seen a tremendous increase in experimental activity.  
High-precision cross section and polarization data are now available, from numerous single and double-meson  
photoproduction experiments, and these have become the broadest source of information on new baryon states. 
The bi-annually released Review of Particle Physics (RPP) is the primary source of information relied upon by
researchers in the field of baryon physics. The RPP tabulates baryon resonance candidates together with 
their properties and provides an assessment of their reliability, both overall and separately from pion and
photon induced reactions. We have therefore focused our discussion on the ratings provided by this source.

The RPP assigns a star-rating from one to four stars for baryon resonance candidates. The one and two-star
states are rated from poor to fair, whereas the three and four-star states have a rating from likely to
certain. These higher-rated states appear in the Baryon Summary Tables without reference to any star rating. 
The more detailed Particle Listings tabulate overall and reaction-specific star ratings for each resonance. 
The three P-wave baryon states, $N(1710)(1/2)^+$, $N(1900){3/2}^+$, and $\Delta(1600){3/2}^+$, are presently
given an overall three-star rating, which prompts the question: Have these states been confirmed or are they
merely 'likely' to exist?

It should be mentioned that neither the definition of a star rating nor the ratings themselves have remained
static since these states were identified. Many states were downgraded, with some three-star rated states
being eliminated, between the 1982~\cite{rpp82} and 1984~\cite{rpp84} editions of the RPP. This upheaval was
prompted by disagreements between newer measurements and the older existing data and analyses~\cite{piN1}.
In particular, the $N(1710)(1/2)^+$ was demoted from a 4-star to a 3-star status, while the $\Delta (1600)(3/2)^+$
dropped from a 3-star to a 2-star status. As a result, the $\Delta (1600)(3/2)^+$ was dropped from the Baryon
Summary Table. Its status was later raised back to three stars~\cite{rpp92}, 
having been seen in subsequent analyses~\cite{VPI85,KSU92}
of the Virginia Tech and Kent State groups, the uncertainty in its properties preventing a 4-star rating.

Many topics related to the extraction of resonance properties, from the 
newly accumulated high-precision data, were subjects of discussions at the 2014 ECT* Workshop 
"Exciting Baryons: Design and Analysis of Complete Experiments for 
Meson Photoproduction" \cite{ECT*Trento}. The presence of many experts in the field enabled a meeting where 
the status of prominent 3-star resonances was discussed, along with possible modifications of the 
RPP~\cite{PDG} star ratings.  One fact was unanimously agreed upon: "While most resonances have been 
discovered by analyzing $\pi N$ elastic scattering data only \cite{Burkert,Anisovich2010}, measuring other channels, 
photoproduction in particular, could confirm some questionable states, and possibly discover new ones." 

It has recently been emphasized that the extraction of resonance information from $\pi N$ elastic amplitudes
requires a detailed knowledge of inelastic channels~\cite{Svarc2013all} as some structures, such as the
$N(1710)$, can be well described either as a pole or a complex branch point, associated with the $\rho N$ channel. 
With only elastic data, there is no clear way to distinguish between the two mechanisms, 
and the inelastic channel information is essential to resolve the 
ambiguity. This fact has only partially been incorporated into the primary analyses of the Karlsruhe-Helsinki 
(KH)~\cite{Hoehler84}, Carnegie-Mellon-Berkeley 
(CMB)~\cite{cmb}, and George Washington University Data Analysis Center (GW DAC)~\cite{gw_pin}, 
and some uncertainties have remained.

The KH analysis assumed that a nearly-unique set of amplitudes could be found for elastic $\pi N$ scattering if
the constraints of isospin invariance and fixed-t analyticity were applied to a set of data with sufficiently
small errors. This claim was supported by a series of mathematical physics studies from Sabba-Stefanescu~\cite{sabba}.
Information from $NN\to \pi\pi$ scattering was also used through dispersion relations. The KH resonance parameters
were initially extracted through a unitary Breit-Wigner plus background fit to the KH amplitudes. In a later
analysis, the speed-plot method was used to determine associated pole parameters. In this way, information
concerning inelastic channels was avoided.

The CMB analysis was similarly focused on the determination of a unique set of $\pi N$ elastic scattering
amplitudes. Constraints were imposed such that the invariant amplitudes were analytic along 5 crossing 
symmetric hyperbolas in the Mandelstam plane. The partial wave analysis used constraints from a Regge
exchange model in which lower partial waves were allowed to vary, with the higher waves fixed. Once determined,
these $\pi N$ partial-wave amplitudes were included in an elaborate multi-channel analysis in order to 
determine resonance parameters. The formalism contained quasi-two-body channels ($\pi \Delta$, $\rho N$, $\eta N$,
$\epsilon N$, $\omega N$, $\pi N*$, $\rho \Delta$), and a non-resonant $\pi \pi N$ channel. These inelastic 
channels were loosely constrained by isobar production cross sections~\cite{cmb_isobar}. Refinements to this
multi-channel approach were subsequently carried out by the Zagreb group~\cite{Zagreb}, emphasizing the 
influence of $\eta N$ production data, and by the Argonne-Pittsburgh group~\cite{Vra00}, which enforced more stringent
constraints from isobar production data. 

The GW DAC analysis was based on a Chew-Mandelstam parameterization of partial-wave amplitudes, building in 
branch cuts for the opening $\pi \Delta$, $\rho N$, and $\eta N$ channels. Forward and fixed-t dispersion
relation constraint were imposed and satisfied interatively in a joint fit to both the constraints and 
the existing experimental data for $\pi N$ elastic scattering and $\eta N$ production data. 

Noticeably absent from the above-described three PWA was the incorporation of existing $KY$ production data. 
In addition, the appearance of
double-polarization data ($R$ and $A$) at at moderate energies~\cite{itep_RA} only after the KH and CMB analyses were completed, has also cast some
doubt on the uniqueness of these elastic amplitudes.

The aim of this Communication is to show that by paying proper attention to inelastic channels we may first 
re-evaluate the status of "old" resonances (resonances which have been extensively discussed for decades), 
and then discuss and establish a level of confidence for newly discovered states.
Photoproduction experiments, particularly  
those producing final-state mesons (such as $K$, $\eta$, $\eta '$, and $\omega$) beyond single
pions~\cite{RBradford,RBradford2007,VCrede,DElsner,MWilliams,McCracken,BDey}   
have been and will continue to be crucial for the discovery of new resonances, 
as has been extensively discussed at ATHOS 2012 Workshop in Camogli 2012 
\cite{Camogli2012} by E. Klempt \cite{Klempt2012}.
\\ \\ \indent
\section{N(1710)1/2$^+$ - N(1710)P$_{11}$}
This resonance is a textbook example showing how the confidence level of a resonance's existence changes through the 
application of different 
analysis techniques, and how our knowledge concerning the necessity of measuring and analyzing inelastic channels has 
matured over time.

As mentioned in the Introduction, this state though originally having a 4-star rating, was demoted to 3 stars in 
the 1984 edition of the RPP. 
The KH and CMB analyses, in spite of fitting mainly
$\pi N$ elastic data, definitely required this state. Some older, and much simpler analyses based 
on the rather small base of inelastic data \cite{Saxon80,Baker79,Baker78,Baker77,Longacre77,Knasel75,Longacre75}
also required this state, so its existence was at least fairly certain. The change in star rating was a result of an 
overall re-evaluation of 
resonance criteria, and notable changes were made for almost all resonances.

Since the 1984 edition, we can separate various PWA attempts in two main categories: i) those that used mainly $\pi N$ 
elastic data, and ii) those that included several, or most of the dominant, inelastic channels.  
Below we summarize the findings of the most well-known efforts.

i)
One of the most elaborate approaches, based almost exclusively on $\pi N$ elastic data was, as described above, performed by KH 
collaboration \cite{Hoehler84}.  In their Breit-Wigner and speed-plot fits to the $P_{11}$ amplitude, they found clear evidence for the N(1710)1/2$^+$ state.
Contrary to this, the GW DAC analysis, also heavily relying on $\pi N$ elastic data, did not find this state in its energy-dependent solution.  
These seemingly contradictory results can, however, be simultaneously understood remembering that, 
as has recently been demonstrated using Laurent+Pietarinen (L+P) technique 
in single-channel analysis~\cite{Svarc2013all}, some of the states with a lower confidence level 
can be alternatively explained either as resonances or as complex branch-points effects due to inelastic channel  
openings ($\rho N$ at W = 1708 - $i$ 70 in particular). There is no way to distinguish 
between the two mechanisms if only the $\pi N$ elastic scattering channel is analyzed.   
As a recent multi-channel analysis from 
the BnGa group finds a very small $\Gamma_{\pi N} / \Gamma$ branching fraction and elastic pole residue, we infer 
that $\pi N$ elastic scattering is not optimal for a determination of its properties. However, 
by using the L+P technique it can be concluded that each of the 
two $\pi N$ analyses is at least consistent with its existence, and no definite conclusion can be made without incorporating inelastic channels. 

The J\"{u}lich coupled-channel model also did not see the N(1710)P$_{11}$ state in their early 
publications \cite{Juelich2003,Ceci2011}. At that stage of their calculation, they had fitted $\pi N$ elastic, and 
somewhat renormalized $\pi N \rightarrow \eta N$ data, and a $\rho N$ complex branch point seemed to be 
sufficient to describe both of these sets of data. However, as we shall see later, when their calculation was 
expanded to include, in particular, strange inelastic channels~\cite{Doering2012}, their conclusions changed. 

ii) The first analysis which included inelastic data more extensively came from Kent State University (KSU)~\cite{Manley1992}, 
using a multichannel approach constrained to fit both $\pi N$ 
elastic and $\pi N \rightarrow \pi \pi N$ data. The existence of N(1710)P$_{11}$ state was unambiguously detected 
because by choosing $\pi \pi N$ channel they were able to keep $\rho N$ complex branch point effects well under 
control. This work was recently updated to include $\eta N$, $K \Lambda$ and single pion photoproduction 
channels \cite{Manley2012}, confirming their earlier conclusions.

The next group of analyses was based on reviving 1979 Cutkosky CMB model, but the data base this time included 
some of the inelastic channels. First, a three-channel model appeared from Zagreb in 1995 \cite{Zagreb} which had 
three coupled channels:  the $\pi N$ elastic channel was represented by KH80 partial waves, the second channel 
contained full available $\pi N \rightarrow \eta N$ data base giving lower confidence to some higher energy data, and 
having the third, effective channel with no data to fit, but left open to allow the transfer of flux from first two channels 
to maintain the multi-channel unitarity.  This model was followed by Argonne-Pittsburgh collaboration 
\cite{Vra00} which extended the number of channels and included additional inelastic channels 
like $\pi \pi N$ for instance. In both cases N(1710)P$_{11}$ was very strongly required. 
The Zagreb group in 2006 also offered the explanation that N(1710)P$_{11}$ state can be explained 
as a continuum ambiguity effect, and that it entirely depends on measuring other channels \cite{Ceci2006}. The conclusion 
that P$_{11}$ state is heavily dominated by inelastic, and in particular $K \Lambda$ channel, 
was later also made by J\"{u}lich group in 2012. In principle, these two statements are equivalent \cite{Ceci2006}.

In following decades many groups developed various forms of coupled-channel models: 
University of Giessen \cite{Giessenall}, EBAC~\cite{Suzuki2010} (presently ANL-Osaka model~\cite{Satoall}), 
Dubna-Mainz-Taipei~\cite{DMT0,DMT1}, 
University of Mainz MAID 2007 \cite{MAID}. They all included various inelastic channels, 
and they all detected the need for N(1710)P$_{11}$ state. 

The J\"{u}lich group also 
included inelastic channels, and instead of just fitting the meager $\pi N$ elastic  
and $\pi N \rightarrow \eta N$ data base, they made a full fit including many other inelastic channels 
($\pi N \rightarrow \eta N$, K$^0 
\Lambda$, K$^0 \Sigma ^0$, K$^+ \Sigma ^-$ and $\pi ^+ p \rightarrow K^+ \Sigma^+$) \cite{Doering2012}. 
Influenced by these inelastic channels, their conclusion about N(1710)P$_{11}$ has changed: 
The state is now definitely needed in their full calculation. 

Finally, the Bonn-Gatchina coupled channel analysis (BnGa) appeared with extremely good results 
for all available pion and photon induced channels \cite{BonnGatchina}. The need for 
N(1710)P$_{11}$ was never in question in 
spite of the presence of many complex branch points which might mask its contribution. 
Here it should be noted that in the single-channel resonance+Regge approach of the Gent group, a fit to $K^+ \Lambda$
photoproduction data~\cite{DeCruz:2011xi} did not require this state. This may, however, be due to a lack of
multi-channel constraints (such as the reaction $\pi N\to K\Lambda$ discussed below) 
for a model very different from those discussed above, and lower coupling of N(1710)P$_{11}$ state to photoproduction channel.

\begin{figure}[h]
\begin{center}
\includegraphics[angle=0,width=0.23\textwidth]{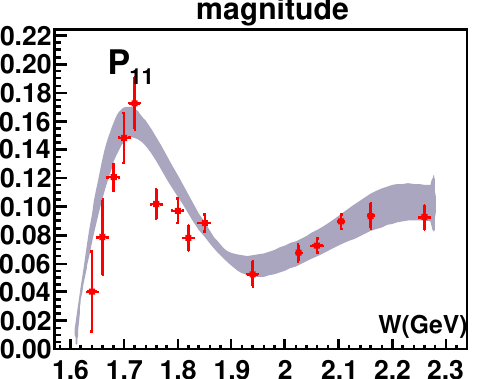}
\includegraphics[angle=0,width=0.23\textwidth]{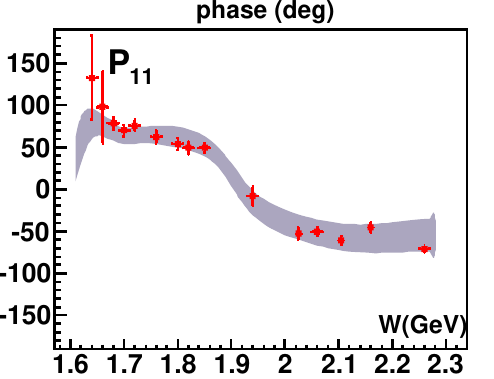}
\caption {(Color online)  Partial wave analysis amplitude (left) and phase shift(right) of 
the inelastic $\pi N \to K\Lambda$ $P_{11}$ wave. The band correspond to a variety of energy-dependent 
BnGa solutions, and points correspond to energy independent solution of ref.~\cite{Anisovich2013}. }
 \label{P11}
\end{center}
\end{figure}

\begin{figure}[h]
\begin{center}
\includegraphics[angle=0,width=0.48\textwidth]{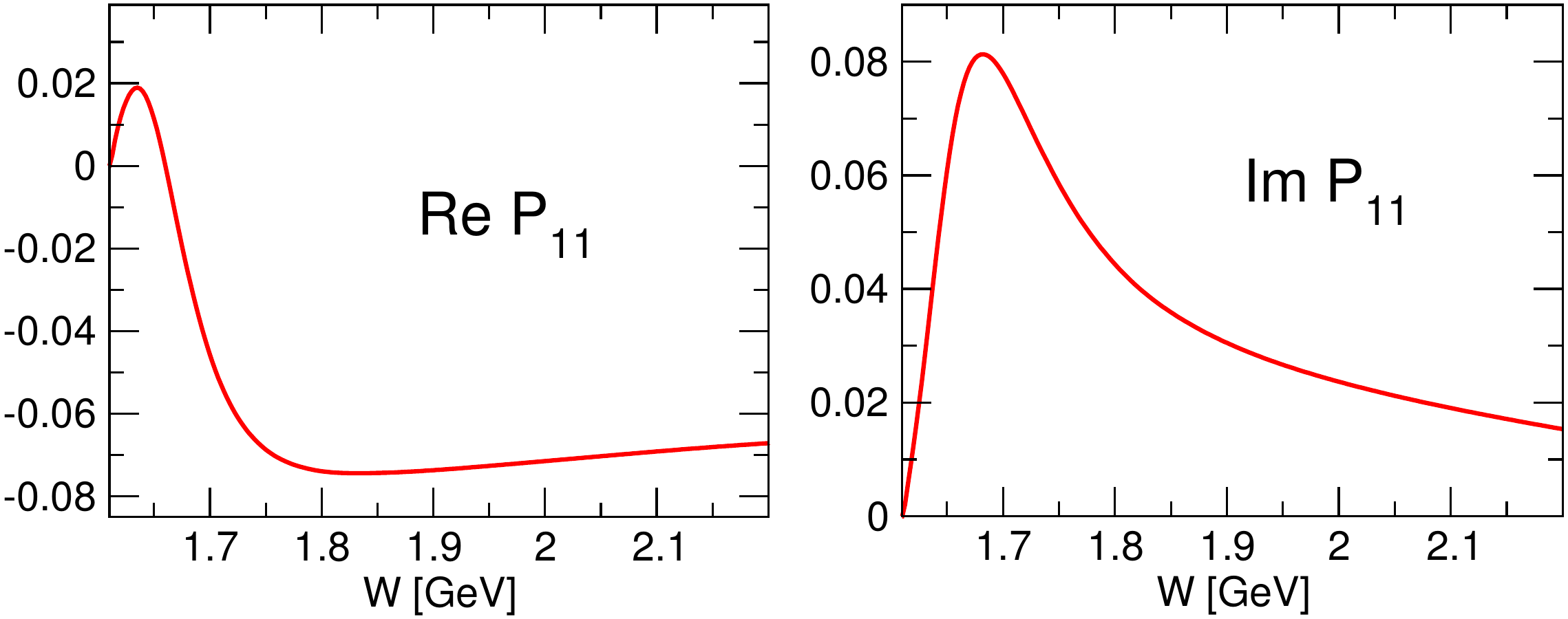}
\caption {(Color online) Real (left) and imaginary part (right) of the $P_{11}$ partial wave amplitude $\tau$  in the inelastic $\pi N \to K\Lambda$ reaction from the J\"ulich2012 model of 
ref.~\cite{Doering2012}.  See ref.~\cite{Doring:2010ap} for a definition of the amplitude $\tau$ and its connection to observable quantities.}
 \label{P11_juelich}
\end{center}
\end{figure}

The analysis of the $\pi N\to K\Lambda$ reaction provides the best example for the importance of data from
inelastic reactions.  First, this reaction is isospin selective: only nucleon states are produced. Second, due to the
self-analyzing power of $\Lambda$ hyperon, the recoil asymmetry can be measured in an unpolarized experiment 
and the rotation parameter in an experiment with the polarized target. Furthermore, in the region around 1700 MeV 
the dominant contributions are expected to come from $S$ and $P$ waves which can be clearly separated in the 
analysis of three observables. Indeed two energy-dependent analyses (J\"ulich~\cite{Doering2012} and BnGa~\cite{Anisovich2013}) found a large contribution 
from the $P_{11}$ wave with a clear resonant structure at 1700 MeV, as shown in Fig.~\ref{P11} for BnGa solution and in Fig.~\ref{P11_juelich} for the J\"ulich solution. This result was confirmed in the 
energy-independent analysis carried out by the Kent and BnGa groups.

Until recently, the reason why different analyses gave conflicting results for the N(1710) P$_{11}$ 
confidence level was not entirely understood, and differences were accepted just as a fact. Thus, there was 
no reason to change the RPP star rating. The natural explanation, that the effects of complex branch-points 
might, in single-channel analyses, be confused with resonance effects, was first considered by the J\"{u}lich - Zagreb 
collaboration~\cite{Ceci2011}, and more transparently explained by the efforts of the Zagreb-Tuzla 
group, by introducing the L+P expansion \cite{Svarc2013all}. 
Therefore, a straightforward conclusion emerged: When a resonance in single-channel analysis can be confused 
with the effects of nearby complex branch-point, channels which are connected with this branch point must also 
be measured to clarify the distinction. 
\\ \\  \noindent
\emph{\underline{Summary}} 
\\ \\ \indent
After realizing that complex branch points may in single-channel analysis give a false resonance signal, 
we can with certainty claim that multi-channel analyses (coupled-channel or combined single-channel ones) 
are essential to unambiguously establish the resonance existence. As all multi-channel analyses have shown 
the need for the N(1710)1/2$^+$ - N(1710)P$_{11}$ state, we believe that its existence has been proven 
beyond any doubt.  
\\ \\ \noindent
\emph{\underline{Recommendation}} 
\\ \\ \indent
Raise the RPP confidence level of N(1710)1/2$^+$ state from 3* to 4*.

\section{$\Delta(1600)3/2 ^+$ - $\Delta (1600)$P$_{33}$}
This is still listed as a 3* resonance, in spite of the fact that its existence is confirmed in each of 
the KH, CMB, and GW DAC
analyses. In addition, a re-analysis of the KH amplitudes, using the L+P method, finds that this cannot
be described by a complex branch-point effect. 

Even the earliest PWA analyses by Longacre \cite{Longacre75,Longacre77} confidently reported its existence. 
A revival of CMB method, extending the fitted data base of inelastic channels, independently by
Zagreb and
Argonne-Pittsburgh \cite{Zagreb,Vra00} similarly find it. 
The same is true for the KSU \cite{Manley1992,Manley2012} and
Giessen \cite{Giessenall}, and ANL-Osaka~\cite{Satoall}. Finally, in the BnGa analysis, it is
present without any doubt \cite{BonnGatchina} in any combination of analyzed data.

While the existence of this state is clear, its mass and width (or pole parameters) are not consistent
between the various determinations. The Breit-Wigner mass, for example, varies from $1510\pm 20$ MeV, in
the BnGa analysis, to 1706$\pm$10 MeV, in the Kent State analysis; 
the width is similarly uncertain, varying from about 200 MeV to
500 MeV in various analyses finding this state.
\\ \\ \noindent
\emph{\underline{Summary}}
\\ \\ \indent
The $\Delta (1600)$P$_{33}$ is confirmed to exist, though its resonance parameters have not been 
accurately determined.
\\ \\ \noindent
\emph{\underline{Recommendation}}
\\ \\ \indent
In spite of the fact that its mass and width are still subject to sizable variations from analysis to analysis, we still
believe that one should seriously consider raising the RPP confidence level of $\Delta (1600)$P$_{33}$ state
from 3* to 4* if this implies existence is certain.

\section{N(1900)3/2$^+$ - N(1900)P$_{13}$} 
This state is a good example of a resonance which appears to couple weakly the to $\pi N$ 
elastic channel. Consequently,
the N(1900)3/2$^+$ was not reported by the KH, CMB, and GW DAC analyses.
However, the KSU analysis \cite{Manley1992}, which had additional constraints from
$\pi \pi N$ channels, did report it; this finding was confirmed in a recent update 
to the KSU analysis~\cite{Manley2012}. Consequently, the RPP has retained the state as a 2* resonance only.
\\ \\ \indent
Using the KH80 partial waves \cite{Hoehler84}, and an analytically smoothed 
version of the same analysis KA84 \cite{Koch85}, the L+P method has unambiguously detected a N(1900)P$_{13}$ state 
in these amplitudes, unreported until now [(1928 $\pm$ 18 $\pm$ 2 ) \mbox{+ $i$ (152 $\pm$ 40 $\pm$ 9)} for KH80 
and (1920 $\pm$ 17 $\pm$ 1) + $i$ (215 $\pm$ 37 $\pm$ 2) for KA84]. Here, however, it was also stressed that this 
structure in the P$_{13}$ partial wave, near 1900 MeV, can alternatively be explained by complex $\rho N$ 
branch-point, so new measurements were needed. 
Therefore, we may say that two "old"  PWA do contain evidence for the N(1900)P$_{13}$ state: 
KSU, which reported it, and KH80, where it remained undiscovered until recently. 
\\ \\ \indent
Other evidence has come from the ANL-Osaka analysis~\cite{Satoall}, 
which included inelastic channels and
reported the $N(1900)P_{13}$, whereas an earlier EBAC analysis~\cite{Suzuki2010} 
of $\pi N$ scattering did not. Another claim comes from the Giessen fit of 2002~\cite{Penner2002} which emphasizes its coupling to the $\omega N$ channel.
\\ \\ \indent
The breakthrough came with new photoproduction experiments. This state was 
solidly established in the 
BnGa coupled-channel analysis \cite{Nikonov2008,Anisovich2012} 
making use of very precise K$\Lambda$ and K$\Sigma$ cross sections and polarization data. 
When the resonance was introduced with varying mass, the $\chi^2$ as a function of the imposed mass showed striking minima for both these reactions \cite{Anisovich:2010an}
and even evidence for its decay into $N\eta$. 
Presently, the reaction $\gamma p \to  p\pi^0\pi^0$ is being studied and it is found 
that the state decays via $\Delta(1232)\pi$ with a branching ratio of about 65\%. In the cascade $N(1900)3/2^+\to N(1520)3/2^-\pi^0\to N\,2\pi^0$, the presence of $N(1900)3/2^+$ can be identified even 
without partial wave analysis when the decay pattern $N^*(1900)\to  N(1520)3/2^-\pi^0\to N\pi^0\pi^0$ from polarized photons is compared to the pattern expected for different $N^*(1900)$ spin-parities 
\cite{Sokhoyan:2014pre}. 
The N(1900)P$_{13}$ was also 
confirmed in an effective Langrangian resonance model 
analysis of $\gamma p \rightarrow K^+ \Lambda$ \cite{Maxwell2012}, 
and in a covariant isobar-model single 
channel analysis of $\gamma p \rightarrow K^+ 
\Lambda$ \cite{Mart2012}.
The discovery of N(1900)P$_{13}$ is a very clear example of the importance of 
the photo-production data for 
establishing the existence of new baryon states. 
Already the $\gamma p\to K\Lambda$ total cross section measured by the CLAS 
collaboration~\cite{RBradford} shows a double-peak structure with the second 
peak at 1900 MeV, indicating 
a contribution from a resonant state (or set of states) with this mass. 
The analysis of the CLAS differential cross section 
and the recoil asymmetry~\cite{RBradford,McCracken} shows 
that this state has 
total spin $J\le 3/2$. 
The determination of the quantum numbers of this state is further clarified from the 
analysis of CLAS data~\cite{RBradford2007} double-polarization observables 
$C_x$ and $C_z$, measured in the 
$\gamma p\to K\Lambda$ and $\gamma p\to K\Sigma$ reactions. 
These data clearly demonstrated that the state, 
which is responsible for the peak in the total cross section, 
has quantum numbers $J^P=\frac 32^+$. 

Finally we want to point out that there are indications in the single channel, event-based analysis of $\omega$ photoproduction
from CLAS~\cite{Williams:2009aa}, and  in the BnGa multi-channel analysis~\cite{Anisovich:2011su}, that there could be 
two N(1900)P$_{13}$ states with close masses. If  confirmed, this could alter properties of the state, but still its existence is unquestionable.  
\\ \\
\emph{\underline{Summary}} 
\\ \\ \indent
The evaluation of this state confirms earlier conclusions. Single channel analysis of $\pi N$ elastic data is difficult, 
and can only give alternative (resonance vs. complex branch point) explanations. Inclusion of some data from 
$\pi\pi N$ channel in KSU analysis \cite{Manley1992} selects the resonance alternative, 
and the CLAS
photoproduction data in the K$^+\Lambda$ channel fully confirmed the state. 
\\ \\ \noindent
\emph{\underline{Recommendation}} 
\\ \\ \indent
Raise the RPP confidence level of N(1900)3/2$^+$ state from 3* to 4*.\\

\section*{Acknowledgement}
We express our gratitude to the European Center for Theoretical Studies in Nuclear Physics and Related 
Areas (ECT$^*$) in Trento who organized the Workshop "Exciting Baryons: Design and Analysis of Complete 
Experiments for Meson Photoproduction" and enabled us to get together and prepare this publication.

\end{document}